\documentclass[10pt,conference]{IEEEtran}
\usepackage[letterpaper, left=0.625in, right=0.625in, bottom=1.05in, top=0.75in]{geometry}
\usepackage{epsfig,setspace,amsmath,epsf,amssymb,bm,theorem,cite,graphicx, epstopdf,algorithm,algpseudocode,float,color,mathtools,authblk,tikz,physics}

\usetikzlibrary{positioning}
\usetikzlibrary{decorations.text}
\usetikzlibrary{decorations.pathmorphing}
\usepackage[short,c2]{optidef}

\usepackage{empheq}
\usepackage{cases}

\usepackage{booktabs}
\usepackage{subfigure}
\usepackage{bbm}

\newtheorem{remark}{Remark}
\newtheorem{lemma}{Lemma}

\newenvironment{Proof}[1]{\medskip\par\noindent{\bf Proof:\,}\,#1}{{\mbox{\,$\blacksquare$}\par}}

\newcommand{\Exp}{\mathrm{Exp}}
\newcommand{\EE}{\mathbb{E}}

\IEEEoverridecommandlockouts
\allowdisplaybreaks

\begin{document}
\title{Dynamic Attack Surfaces: A Queue of \\ Branching Processes}

\author[1]{Ismail Cosandal}
\author[2]{Abdullah Y. Etcibasi}
\author[2]{C. Emre Koksal}
\author[1]{Sennur Ulukus}

\affil[1]{\normalsize University of Maryland, College Park, MD, USA}
\affil[2]{\normalsize The Ohio State University, Columbus, OH, USA}
\maketitle

\begin{abstract}
Modern attack surfaces are inherently dynamic. Vulnerabilities emerge over time and they remain exposed to exploit until defenders discover and patch them. Artificial intelligence (AI) amplifies these dynamics by accelerating both offensive and defensive processes, making modern cybersecurity increasingly a competition among interacting time scales and resources. We capture this competition using a queue of branching processes that couples vulnerability discovery and resource-constrained patching with the evolving attacker population targeting each vulnerability. Our primary risk metric is the cumulative number of exploits a vulnerability experiences before it is patched.  We derive the expected cumulative number of exploits under birth-death-immigration (BDI) and rise-decay (RD) models of attacker activity. Under a BDI model of attacker activity, we identify a phase transition: the expected cumulative number of exploits is finite only when attacker growth remains below both the vulnerability discovery rate and the effective patching rate. Across both attacker models, increasing the exploit rate does not necessarily increase the expected number of exploits because faster attacker departure can sufficiently limit the active attacker population. Finally, we show that equal proportional AI amplification of offensive and defensive capabilities need not preserve security: accelerating both sides can shift the defensive bottleneck and move the system into an unbounded-risk regime.
\end{abstract}

\section{Introduction}
The growing use of artificial intelligence (AI) by both attackers and defenders underscores the need for a dynamic model of the attack surface. AI agents can continually probe systems, discover and exploit weaknesses, and amplify attacker activity. At the same time, defenders can use AI to accelerate vulnerability discovery, prioritization, and patch generation. The resulting attack surface is therefore not a fixed inventory of flaws, but an evolving system governed by competing offensive and defensive processes \cite{etcibasi2026resource, yun2026dynamic}. This acceleration is already observed in practice. In a reported 2026 campaign against recently disclosed PaperCut vulnerabilities, a threat actor used hundreds of AI agents to develop, test, and execute exploits, reaching the first real-world remote code execution in under four hours and compromising at least 440 exposed instances across 48 countries \cite{french2026papercut}. Static quantities such as vulnerability counts, severity scores, or average patch times cannot capture this temporal dynamics \cite{jones2011fair}. 

To that end, we model the attack surface as a dynamic set of active vulnerabilities that attract evolving attacker populations until patched. Since vulnerabilities arrive over time, compete for limited defensive resources, and depart after patching, queueing theory naturally captures their lifetimes and the competition between attack and defense. Recent work modeled the attack surface as a vulnerability queue and showed that underlying defensive resources, including effective workforce size and aggregate patching capacity, can be reconstructed from vulnerability discovery and patching timestamps \cite{etcibasi2026resource}. Here, we further establish that cyber risk is shaped not only by the rate at which vulnerabilities emerge, but also by congestion and resource limitations within the defensive workflow. Our novel model captures the potentially growing population of attackers targeting each vulnerability over its unpatched lifetime.

Specifically, motivated by the rate competition between the attack and the defense, we introduce a \emph{queue of branching processes}. While a vulnerability remains unpatched, the population of attackers targeting it evolves as a branching process. We consider both a birth-death-immigration (BDI) model, which captures persistent arrival and attacker-to-attacker amplification, and a rise-decay (RD) model, which captures an initial period of rapid attention followed by diminishing interest. The resulting framework couples two time scales that are commonly analyzed separately: the random lifetime of a vulnerability under resource-constrained patching and the growth of its associated attacker population.

Under the BDI model of attacker activity, the central result reveals a phase transition in the expected cumulative number of exploits: the expected number of exploits remains finite only when attacker activity grows more slowly than both vulnerability discovery and the effective patching process. If attacker activity surpasses either defensive mechanism, the expected cumulative number of exploits becomes unbounded. This does not imply that every vulnerability generates infinitely many exploits. Instead, vulnerabilities that experience unusually long discovery or patching delays may attract exceptionally high exploit activity, causing the average number of exploits across all vulnerabilities to diverge. For both models, vulnerabilities attract new attackers at a rate that depends on the number of active attackers, while each active attacker exploits the vulnerability at a constant rate.

An important observation is that the expected number of exploits is not necessarily monotone with the exploit rate. In particular, increasing the exploit rate may reduce the active attacker population sufficiently to offset the direct increase in exploit activity, resulting in a non-monotone relationship. The result therefore identifies vulnerability discovery and patching capacity as two distinct defensive bottlenecks. Strengthening only one is insufficient when the other remains unable to keep pace with attacker growth.

To illustrate the transition in the context of AI-amplified attack and defense, we provide an insightful numerical example (see details in Section ~\ref{sec:numerical_results}) in which we scale the AI capabilities for both offensive and defensive operations by the same factor. We observe that these proportional improvements in attack and defense capabilities do not necessarily preserve security. In this scenario, although patching capacity improves sufficiently, vulnerability discovery becomes the limiting defensive mechanism and is overtaken by attacker growth. Therefore, defensive AI must strengthen both discovery and patching sufficiently to remain ahead of the growth in AI-enabled attacker activity.

The main contributions of this paper are as follows:
\begin{itemize}
    \item To model the dynamics of an attack surface, we introduce a branching processes framework that couples vulnerability discovery and resource-constrained patching with the stochastic evolution of the attacker population targeting each vulnerability.

    \item We derive vulnerability lifetime distribution and closed-form expressions for the expected cumulative number of exploits under a BDI attack model.

    \item We establish a stability condition and show that violating this condition yields infinite expected number of exploits.

    \item We study the allocation of defensive resources between vulnerability discovery and patching and extend it to an RD model in which attacker interest eventually vanishes.
\end{itemize}

\section{System Model} \label{sec:sys}
\begin{figure}
    \centering
    \includegraphics[width=0.85\linewidth]{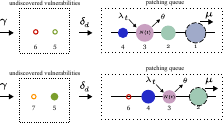}
    \caption{The queue model. We illustrate the state of the vulnerability queue at two separate time instances. The size of each filled circle denotes the number of active attackers, which is governed by $\lambda_t$ and $\theta$, and an empty circle means the vulnerability has not been discovered by any attacker yet.}
    \label{fig:queue}
    \vspace*{-0.4cm}
\end{figure}

We consider a cybersecurity system, illustrated in Fig.~\ref{fig:queue},
where system vulnerabilities are created as a Poisson process with arrival rate $\gamma$. For each existing vulnerability, attackers and the server compete with each other. The goal of the server is to patch the vulnerabilities before the vulnerability attracts many new attackers, and they exploit it. Each existing vulnerability is discovered by the defender after a random time $X$, which is exponentially distributed with rate $\delta_d$; we denote it by $X\sim\Exp(\delta_d)$. We consider the first-discovered-first-served principle, analogous to the first-come-first-served (FCFS) principle, in which each discovered vulnerability is added to a queue, and patching lasts $S\sim\Exp(\mu)$. 

Similar to the defender side, a single attacker discovers an existing vulnerability after $Y\sim\Exp(\delta_a)$. We define a branching process $N(t)$ that represents the active attackers who are attempting to exploit the vulnerability, and $t\geq0$ is the time elapsed after the creation of the vulnerability. The branching process is initiated with $N(0)=0$ and remains zero until an attacker discovers it at $Y$. Each vulnerability attracts new attackers with attraction intensity $\lambda_t$ at time $t$. In addition, each active attacker exploits the vulnerability individually at a constant rate $\theta$, then leaves. We consider two branching processes, BDI and RD, whose attraction intensities are 
\begin{align}
    \lambda_t^{\text{bdi}}&=\lambda N(t)+\lambda_{0},\\
    \lambda_t^{\text{rd}}&=\lambda e^{-\beta(t-Y)}N(t).
\end{align}
In the BDI process, attacker attraction increases with the active attacker population with an additional immigration rate $\lambda_{0}$, whereas in the RD process, the attraction rate associated with each active attacker decays after attacker discovery. Further details of these processes are provided in Section~\ref{sec:analysis}.

We are interested in the number of exploits by the attackers before the patching of the vulnerability. We denote the expected number of active attackers at time $t$ as
\begin{align}
    n(t)=\EE[N(t)].
\end{align}
Since each active attacker exploits the vulnerability at a constant rate $\theta$, the expected number of exploits can be expressed as the expected value of the following integral
\begin{align}
    \bar{D}=\EE\left[\int_0^T \theta n(t)\mathrm{d}t\right], \label{eq:barD_def} 
\end{align}
where $T$ is the \emph{lifespan of a vulnerability}, corresponding to the time from creation until patch completion. In Section~\ref{sec:analysis}, we show that $T$ has a steady-state distribution in a stable queue that does not depend on the creation order of vulnerabilities and is independent of the branching process $N(t)$.

The patching queue model is illustrated in Fig.~\ref{fig:queue} for two different time instants. Each circle represents a vulnerability, and they are enumerated in their creation order. The size of the circles denotes the number of active attackers attacking the vulnerability, and an empty circle means that there are no active attackers using the vulnerability. In the first time instance, a total of six vulnerabilities exist; while four of them are in the patching queue, the remaining two vulnerabilities are not discovered by the defender or attackers yet. Since the discovery durations of vulnerabilities are independent of each other, the $6$th vulnerability is discovered by the defender before the $5$th one in this example. Additionally, the $5$th vulnerability has been discovered by the attackers before it is added to the patching queue. Since the defender and attacker discovery durations are independent exponential random variables, the probabilities that a vulnerability is first discovered by the defender or by an attacker are $\frac{\delta_d}{\delta_d+\delta_a}$ and $\frac{\delta_a}{\delta_d+\delta_a}$, respectively.

\section{Analysis} \label{sec:analysis}
In this section, we first analyze the queue model to obtain the distribution of a vulnerability's lifespan, from its creation until patch completion. Then, we analyze two types of branching processes, BDI and RD, to obtain the expected number of exploits under different scenarios.

\subsection{Queue Analysis}
We consider $\gamma<\mu$, under which the number of unpatched vulnerabilities does not accumulate without bound. The complementary case $\gamma\geq\mu$ is a trivial case where the queue length grows without bound. Each vulnerability is added to the patching queue of vulnerabilities after a random time $X$ following its creation. From the displacement theorem \cite{baccelli2010stochastic}, vulnerabilities arrive at the patching queue with rate $\gamma$, since $X_i$ are i.i.d.~discovery durations. In other words, the patching queue is an $M/M/1$ queue with arrival rate $\gamma$ and service rate $\mu$, and uses the FCFS principle. 

Let us denote the lifespan of the $i$th vulnerability by $T_i$, which can be expressed as 
\begin{align}
    T_i=X_i+W_i+S_i, \label{eq:Ti}
\end{align}
where $X_i$, $W_i$ and $S_i$ are the discovery duration, waiting duration in the queue and the patching duration of the vulnerability $i$, respectively. A well-known property of an $M/M/1$ queue with the arrival rate $\gamma$ and service rate $\mu$ is that the sum of a packet's waiting and service durations follows an exponential distribution with rate $\mu-\gamma$ at steady state under the stability condition, i.e., $\gamma<\mu$ \cite{bolch2006queueing}. Consequently, in the steady state, each $T_i$ has the same distribution, with density denoted by $f_T(t)$, that can be obtained for $\delta_d\neq\mu-\gamma$ as
\begin{align}
    f_T(t)= \frac{\delta_d(\mu-\gamma)}{\delta_d-\mu+\gamma}\left(e^{-(\mu-\gamma)t}-e^{-\delta_d t}\right),  \label{eq:pdfT}
\end{align}
which is the distribution of sum of two independent exponential random variables with different parameters. For the case of $\delta_d=\mu-\gamma$, the distribution becomes an Erlang-$2$ distribution, which is the distribution of sum of two exponential random variables with identical parameters: $f_T(t)=\delta_d^2te^{-\delta_d t}$. Since the function $f_T(t)$ in \eqref{eq:pdfT} is continuous in $\delta_d$, Erlang-2 distribution can be obtained as the limit of \eqref{eq:pdfT}, and therefore, in the rest of the paper, our analysis will focus on the case $\delta_d\neq\mu-\gamma$, with the equality case obtained by taking the limit.

\begin{lemma}
    Under a defense constraint $\mu+\delta_d=b$, $\bar{D}$ is minimized for any branching process independent of the vulnerability lifespan with the choice of $\delta_d=\frac{b-\gamma}{2}$ and $\mu=\frac{b+\gamma}{2}$ under the condition $b>\gamma$. \label{lem:st}
\end{lemma}

\begin{Proof}
First, as shown in \cite{boland1994schur}, among convolutions of independent exponential random variables whose rates have a fixed sum, greater dispersion of the rate parameters results in a larger random variable in the likelihood-ratio order. Let us define $\alpha=\mu-\gamma$; then the budget constraint becomes $\alpha+\delta_d=b-\gamma$. Notice that among all feasible parameters $(\alpha,\delta_d)$, the least discrepancy is achieved with balanced parameters, i.e., $\alpha^*=\delta_d^*=\frac{b-\gamma}{2}$, and the condition $b>\gamma$ guarantees that these rates are positive.

Next, we define $T^*$ as the vulnerability lifespan under the balanced parameters $\alpha^*$ and $\delta_d^*$. Since the likelihood-ratio order implies the usual stochastic order \cite{shaked2007stochastic}, the above result gives
\begin{align}
    T^* \leq_{\mathrm{st}} T
\end{align}
for every other feasible allocation $T$, where $\leq_{\mathrm{st}}$ denotes the usual stochastic order. In addition, the usual stochastic order implies the same ordering of the expectations of non-decreasing functions, whenever the corresponding expectations exist. Notice that the accumulated quantity $\int_0^t \theta n(s)\,\mathrm{d}s$ in \eqref{eq:barD_def} is a non-decreasing function of $t$, since $n(t)\geq0$. Therefore, the usual stochastic order gives
\begin{align}
    \EE\left[ \int_0^{T^*} \theta n(s)\,\mathrm{d}s   \right]
    \leq
    \EE\left[ \int_0^T \theta n(s)\,\mathrm{d}s\right],
\end{align}
which completes the proof. 
\end{Proof}

\subsection{Birth-Death-Immigration (BDI) Model}
Here, we investigate the BDI process with attraction intensity $\lambda_t=\lambda N(t)+\lambda_{0}$. In this model, each active attacker attracts a new attacker at a rate $\lambda$, and new attackers are independently attracted to the vulnerability at a rate $\lambda_{0}$. For simplicity of the analysis, we assume that the immigration rate and the attacker discovery rate are equal, i.e., $\delta_a=\lambda_{0}$. Under this assumption, the first immigrant represents the attacker whose discovery duration is $Y$. The ordinary differential equation (ODE) for $n(t)$ can be written as
\begin{align}
    \frac{\mathrm{d}n(t)}{\mathrm{d}t}=\lambda n(t)+\lambda_{0}-\theta n(t), \label{eq:ni}
\end{align}
with initial condition $n(0)=0$. Hence, the expected number of active attackers at time $t$ can be found for $\theta\neq\lambda$ as
\begin{align}
    n(t)=\frac{\lambda_{0}}{\kappa}\left(e^{\kappa t}-1\right),
\end{align}
where we define $\kappa=\lambda-\theta$ as the attack growth rate. Notice that this expression is continuous at $\kappa=0$, since $\lim_{\kappa\to 0} n(t)=\lambda_{0} t$, which describes the population growth when the attraction and exploit rates are equal. Since the case $\kappa=0$ is recovered by continuity, we consider only $\kappa\neq0$ for the rest of the analysis. In addition, we define the expected number of exploits up to time $t$ by $D(t)$, which is given as
\begin{align}
    D(t)=\int_0^t\theta n(s)\mathrm{d}s=\frac{\theta \lambda_{0}}{\kappa}\left(\frac{e^{\kappa t}-1}{\kappa}-t\right). \label{eq:Dt}
\end{align}

We can now obtain the expected number of exploits before patch completion by averaging $D(t)$ over the lifespan distribution in \eqref{eq:pdfT} as
\begin{align}
   \bar{D}&=\int_0^\infty D(t)f_T(t)\mathrm{d}t \nonumber \\
     &= \dfrac{\delta_d(\mu-\gamma)}{\delta_d-\mu+\gamma}\left[\int_0^\infty D(t) e^{-(\mu-\gamma) t}\mathrm{d}t-\int_0^\infty D(t) e^{-\delta_d t} \mathrm{d}t\right]. \label{eq:bar2}
\end{align}
Notice that this expression applies to any branching process for which $D(t)$ is well-defined and independent of the vulnerability lifespan. Now, defining the unilateral Laplace transform of the function $D(t)$ as
\begin{align}
    \mathcal{L}\{D(t)\}(s)=\int_0^\infty D(t) e^{-s t} \mathrm{d}t,
\end{align}
the expression in \eqref{eq:bar2} for $\bar{D}$ can be written as
\begin{align}
    \bar{D}=\frac{\delta_d(\mu-\gamma)}{\delta_d-\mu+\gamma}\Big(\mathcal{L}\{D(t)\}(\mu-\gamma)-\mathcal{L}\{D(t)\}(\delta_d)\Big)
\end{align}
Using \eqref{eq:Dt}, $\mathcal{L}\{D(t)\}(s)$ can be found as
\begin{align}
    \mathcal{L}\{D(t)\}(s) = \dfrac{\theta \lambda_{0}}{\kappa} \left( \frac{1}{\kappa} \mathcal{L}\{e^{\kappa t}\}(s) - \frac{1}{\kappa} \mathcal{L}\{1\}(s) - \mathcal{L}\{t\}(s) \right),
\end{align}
where $\mathcal{L}\{1\}(s) = \frac{1}{s}$, $\mathcal{L}\{t\}(s) = \frac{1}{s^2}$ for $s>0$ \cite{oppenheim1997signals}. In addition, $\mathcal L\{e^{\kappa t}\}(s)=\frac{1}{s-\kappa}$ for $s>\kappa$, while the transform diverges for $s\leq\kappa$. Now, we can obtain the closed-form expression for $\bar{D}$ by rearranging terms, and defining $g=\frac{\delta_d(\mu-\gamma)}{(\mu-\gamma-\kappa)(\delta_d-\kappa)}$, as
\begin{align}
     \bar{D}&= \begin{cases}
             \dfrac{\theta\lambda_{0}}{\kappa}\left(\dfrac{g-1}{\kappa}-\dfrac{1}{\mu-\gamma}-\dfrac{1}{\delta_d}\right) , & \min(\delta_d,\mu-\gamma)>\kappa, \\
            \infty, & \min(\delta_d,\mu-\gamma)\leq\kappa,
        \end{cases} \label{eq:barD}
\end{align}
where we recall that $\kappa=\lambda-\theta$.

We note that \eqref{eq:barD} establishes the central phase transition of the BDI model and connects it to AI-amplified cyber risk. Offensive AI can push attacker growth beyond the critical boundary, while defensive AI can accelerate vulnerability discovery and patching. The expected cumulative number of exploits remains finite only when both defensive mechanisms stay ahead of the attacker growth; that is, we must have both $\delta_d>\kappa$ and $\mu-\gamma>\kappa$. Otherwise, rare vulnerabilities with long discovery or patching delays can make the expectation unbounded. Thus, defensive AI must strengthen both mechanisms (vulnerability discovery and patching) rather than only the faster one.

\begin{remark}
    Let us define $\lambda_c=\frac{\delta_d(\mu-\gamma)}{\delta_d+\mu-\gamma}$. The behavior of $\bar{D}$ with respect to $\theta$, denoted by $\bar{D}(\theta)$, can be summarized as follows: i) $\bar{D}(\theta)$ is strictly increasing for $0<\lambda\leq\lambda_c$, ii) $\bar{D}(\theta)$ has a unique maximum for $\lambda_c<\lambda<\min(\mu-\gamma,\delta_d)$, iii) $\bar{D}(\theta)$ is strictly decreasing for $\min(\mu-\gamma,\delta_d)\leq\lambda$ (except $\theta\leq\lambda-\min(\mu-\gamma,\delta_d)$ where $\bar{D}(\theta)$ diverges).  \label{rem:bdi}
\end{remark}

Remark~\ref{rem:bdi} follows by analyzing the sign of the derivative of $\bar D(\theta)$, and the proof is omitted due to space limitations.

\subsection{Rise-Decay (RD) Model}
In the BDI model of the previous subsection, the expected attacker population can grow exponentially without bound when $\lambda>\theta$. Consequently, the expected number of exploits may diverge when the discovery and patching resources of the defender are insufficient relative to this growth. In this subsection, we consider the RD process as an alternative, where the expected number of active attackers can initially increase, while the attraction rate gradually diminishes. Since the attacker population is initiated by the first discovery at $t=Y$, we assume that the attraction rate associated with each active attacker decays from this epoch. Specifically, we express the attraction intensity for this model as
\begin{align}
    \lambda_t=\lambda e^{-\beta (t-Y)}N(t), \quad t\geq Y,
\end{align}
where $\beta$ determines the exponential decay of attacker attraction.

As defined in Section~\ref{sec:sys}, the number of active attackers is zero until the vulnerability is first discovered at $t=Y$, at which point size of the  population becomes one. We therefore shift the time origin to the discovery epoch by defining $\sigma=t-Y$ as the elapsed duration since the first attacker's discovery. For notational simplicity, we continue to use $n(\sigma)$ for the expected number of active attackers after this shift. Then, the ODE for $n(\sigma)$ can be expressed as
\begin{align}
    \dfrac{\mathrm{d}n(\sigma)}{\mathrm{d}\sigma}=\lambda e^{-\beta \sigma} n(\sigma)-\theta n(\sigma), \quad \sigma\geq 0, \label{eq:ni_2}
\end{align}
with initial condition $n(0)=1$. The resulting $n(\sigma)$ and $D(\sigma)$ can be expressed as
\begin{align}
    n(\sigma)=&\exp\left(\frac{\lambda}{\beta}-\frac{\lambda}{\beta}e^{-\beta \sigma}-\theta \sigma\right),\\
    D(\sigma)=&\theta\int_0^\sigma \exp\left(\frac{\lambda}{\beta}-\frac{\lambda}{\beta}e^{-\beta s}-\theta s\right) \mathrm{d}s. \label{Dsigma}
\end{align}
In addition, we define $\Sigma=\max(0, T-Y)$, which represents the remaining vulnerability lifespan after attacker's discovery when the discovery occurs before patch completion, and is zero otherwise. The density of $\Sigma$ for $\sigma>0$, can be obtained by
\begin{align}
    f_{\Sigma}(\sigma)=&\int_0^\infty f_T(\sigma+s) \delta_ae^{-\delta_as}\mathrm{d}s \nonumber \\
    =&k_0
    \left(\frac{1}{k_1}e^{-(\mu-\gamma)\sigma}
    -\frac{1}{k_2}e^{-\delta_d \sigma}\right), \label{fSigma}
\end{align}
where $k_0=\frac{\delta_d\delta_a(\mu-\gamma)}
{\delta_d-\mu+\gamma}$, $k_1=\mu-\gamma+\delta_a$ and $k_2=\delta_d+\delta_a$. Note that the distribution of $\Sigma$ has a mass at zero,  $\mathbb{P}(\Sigma=0)=\frac{\delta_d(\mu-\gamma)}{(\delta_d+\delta_a)(\mu-\gamma+\delta_a)}$. However, since $D(0)=0$, the atom at $\Sigma=0$ makes no contribution to the expected number of exploits. 

\begin{figure*}[t]
    \centering
    \begin{minipage}[t]{0.27\textwidth}
        \centering
        \includegraphics[width=\linewidth]{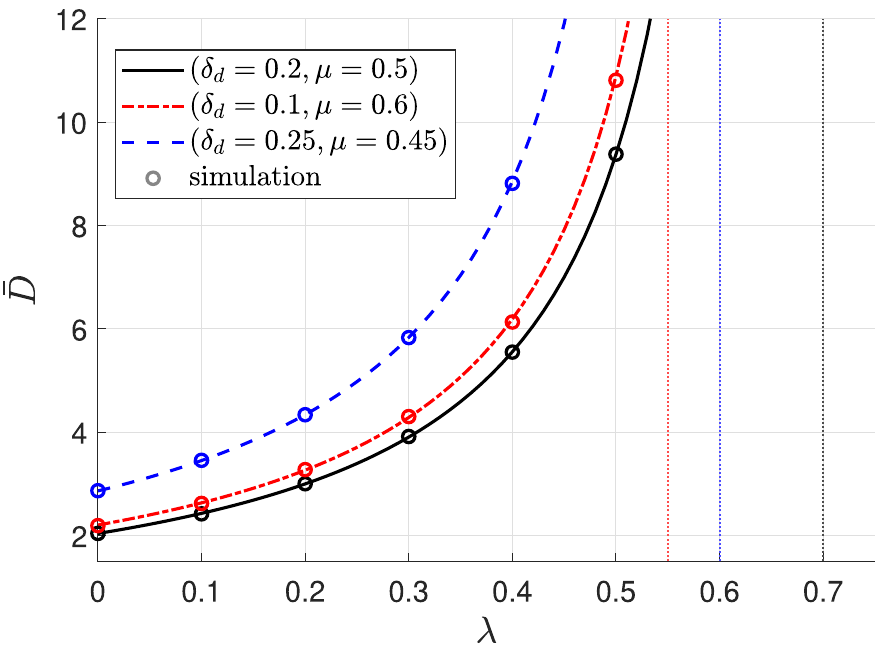}
    \end{minipage}
    \hfill
    \begin{minipage}[t]{0.27\textwidth}
        \centering
        \includegraphics[width=\linewidth]{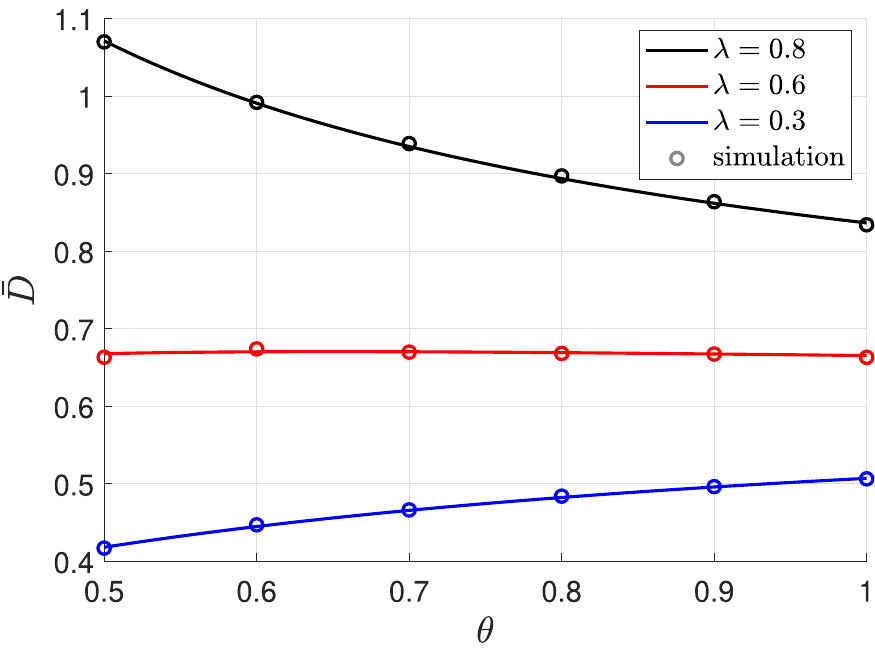}
    \end{minipage}
    \hfill
    \begin{minipage}[t]{0.27\textwidth}
        \centering
        \includegraphics[width=\linewidth]{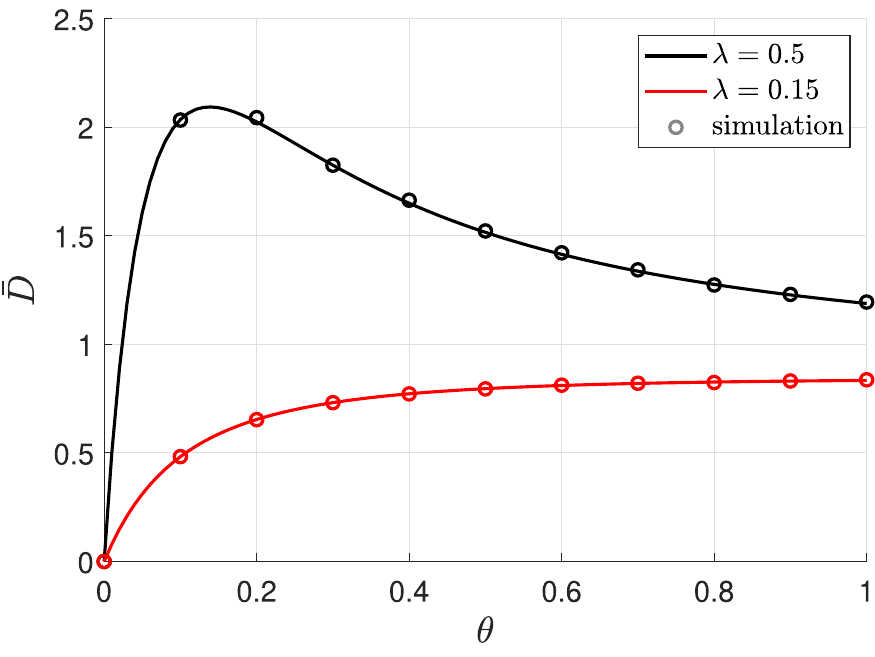}
    \end{minipage}
    \caption{The average exploit of a vulnerability $\bar{D}$ versus: (a) the attraction rate $\lambda$ for the BDI model, (b) the exploit rate $\theta$ for the BDI model, (c) the exploit rate $\theta$ for the RD model.}
    \label{fig:bdi-rd}
    \vspace*{-0.4cm}
\end{figure*}

In the RD model, $\bar{D}$ can be expressed in terms of the unilateral Laplace transform using \eqref{fSigma} as
\begin{align}
    \bar{D}&=\int_0^\infty D(\sigma)f_{\Sigma}(\sigma)\mathrm{d}\sigma \nonumber\\
    &=k_0\left(\frac{1}{k_1}\mathcal{L}\{D(\sigma)\}(\mu-\gamma)-\frac{1}{k_2}\mathcal{L}\{D(\sigma)\}(\delta_d)\right).
\end{align}
Using \eqref{Dsigma}, the unilateral Laplace transform $\mathcal{L}\{D(\sigma)\}(a)$ is
\begin{align}
    \!\!\mathcal{L}\{D(\sigma)\}(a)&=\theta\int_0^\infty \int_0^\sigma \exp(\frac{\lambda}{\beta}-\frac{\lambda}{\beta}e^{-\beta s}-\theta s)\mathrm{d}s e^{-a\sigma}\mathrm{d}\sigma \nonumber\\ 
    &=\dfrac{\theta}{a} \underbrace{\int_0^\infty \exp(\frac{\lambda}{\beta}-\frac{\lambda}{\beta}e^{-\beta s}-(\theta+a) s)\mathrm{d}s.}_{J(\theta+a)} \!\!
\end{align}
By change of variables $x=e^{-\beta s}$, the function $J(y)$ becomes
\begin{align}
    J(y)=&\int_0^\infty \exp(\frac{\lambda}{\beta}-\frac{\lambda}{\beta}e^{-\beta s}-y s)\mathrm{d}s \nonumber \\
    =& \dfrac{e^{\frac{\lambda}{\beta}}}{y}{}_1\!F_1\left(\frac{y}{\beta }; 1+\frac{y}{\beta }; -\frac{\lambda}{\beta}\right),
\end{align}
where ${}_1\!F_1(a; b;c)$ is the confluent hypergeometric function \cite{abramowitz1948handbook}.

\begin{remark}
    Similar to the case of the BDI model, $\bar{D}(\theta)$ in the RD model here has the following properties depending on the value of $\lambda$: i) $\bar{D}(\theta)$ is strictly increasing for $\lambda\leq \lambda_c$, ii) $\bar{D}(\theta)$ has a unique maximum for $\lambda> \lambda_c$, where $\lambda_c=\frac{\delta_d(\mu-\gamma)}{\delta_d+\delta_a+\mu-\gamma}$. \label{rem:rd}
\end{remark}

Remark~\ref{rem:rd} follows by comparing the per-attacker attraction rate $\lambda e^{-\beta \sigma}$ with the ratio $\frac{f_\Sigma(\sigma)}{\mathbb P(\Sigma>\sigma)}$. The former decreases with $\sigma$, while the latter increases, and their values at $\sigma=0^+$ are $\lambda$ and $\lambda_c$, respectively. This comparison determines the sign of the derivative of $\bar D(\theta)$ and whether it changes. The remaining algebra is omitted due to space limitations.

\section{Numerical Results} \label{sec:numerical_results}
To verify our analysis, we run simulations under different parameter settings and branching process models. In each simulation, at least $10^6$ vulnerabilities are generated, and the first $10^4$ vulnerabilities are excluded when computing the average number of exploits to allow the queue to reach steady state. For the RD model, the next event time corresponding to the arrival of a new attacker is simulated using the inverse-integrated-intensity method, which follows from the time-rescaling theorem \cite{brown2002time}. 

Fig.~\ref{fig:bdi-rd} illustrates the simulation results under the following three settings. In these figures, curves represent the analytical results from Section~\ref{sec:analysis}, and simulation results are denoted by circles. In the first setting, we consider a BDI model with fixed parameters $\gamma=0.3$, $\theta=0.2$, $\lambda_{0}=\delta_a=0.5$ and illustrate $\bar{D}$ for varying $\lambda$ values and three $(\mu,\delta_d)$ pairs that satisfy a defense resource constraint $\mu+\delta_d=b=0.7$. Curves in Fig.~\ref{fig:bdi-rd}(a) are obtained from \eqref{eq:barD}, and corresponding simulation results are denoted by circles. The match between analytical curves and simulation results verifies the analysis. In this setting, $\bar{D}$ is finite in the region $\lambda<\min(\delta_d,\mu-\gamma)+\theta$, and the corresponding boundaries for the three parameter pairs are represented by vertical lines. Notice that $\bar{D}$ is minimized for any $\lambda$ when the defense resources are chosen as $\mu=\frac{b+\gamma}{2}=0.5$ and $\delta_d=\frac{b-\gamma}{2}=0.2$, verifying Lemma~\ref{lem:st}. In addition, this allocation maximizes $\min(\delta_d,\mu-\gamma)$, and thus, the range of $\lambda$ for which $\bar{D}$ remains finite. Thus, balancing the defense resources according to Lemma~\ref{lem:st} not only minimizes the expected number of exploits but also allows the defense system to withstand higher attacker attraction rates without divergence.

In the second setting, we verify Remark~\ref{rem:bdi} for the BDI model. We fix the parameters $\gamma=0.3$, $\delta_d=1$, $\lambda_{0}=\delta_a=0.25$, and $\mu=1$, yielding $\lambda_c=0.41$ and $\min(\delta_d,\mu-\gamma)=0.7$. Fig.~\ref{fig:bdi-rd}(b) illustrates $\bar{D}$ for varying $\theta$ with three different $\lambda$ values. We observe that when $\lambda=0.3<\lambda_c$, the direct effect of increasing the exploit rate dominates the reduction in the active attacker population, and hence $\bar D$ increases with $\theta$. For $\lambda_c<\lambda=0.6<\min(\delta_d,\mu-\gamma)$, these two effects dominate in different ranges of $\theta$, resulting in a unique maximum. When $\lambda=0.8>\min(\delta_d,\mu-\gamma)$, increasing $\theta$ suppresses the growth of the attacker population sufficiently strongly that $\bar D$ decreases.

We next repeat a similar experiment for the RD model with parameters $\gamma=0.3$, $\delta_d=0.5$, $\delta_a=0.5$, $\mu=1$ and $\beta=0.1$ in Fig.~\ref{fig:bdi-rd}(c). These parameters yield $\lambda_c=0.21$ in Remark~\ref{rem:rd}, and we consider two values of $\lambda$, $\lambda=0.5>\lambda_c$ and $\lambda=0.15<\lambda_c$, for varying $\theta$. Consistent with Remark~\ref{rem:rd}, we observe that when $\lambda=0.15<\lambda_c$, the reduction in the active attacker population caused by increasing $\theta$ is not sufficient to offset the higher exploit rate, and hence $\bar D$ increases with $\theta$. In the case $\lambda=0.5>\lambda_c$, although increasing $\theta$ initially increases the exploit rate, an excessively large $\theta$ causes active attackers to leave too quickly for the attacker population to grow sufficiently, which reduces the expected number of exploits. The close match between the analytical and simulation results also verifies the analysis for the RD model.

\begin{figure}[!t]
	\centering
	\includegraphics[width=0.33\textwidth]{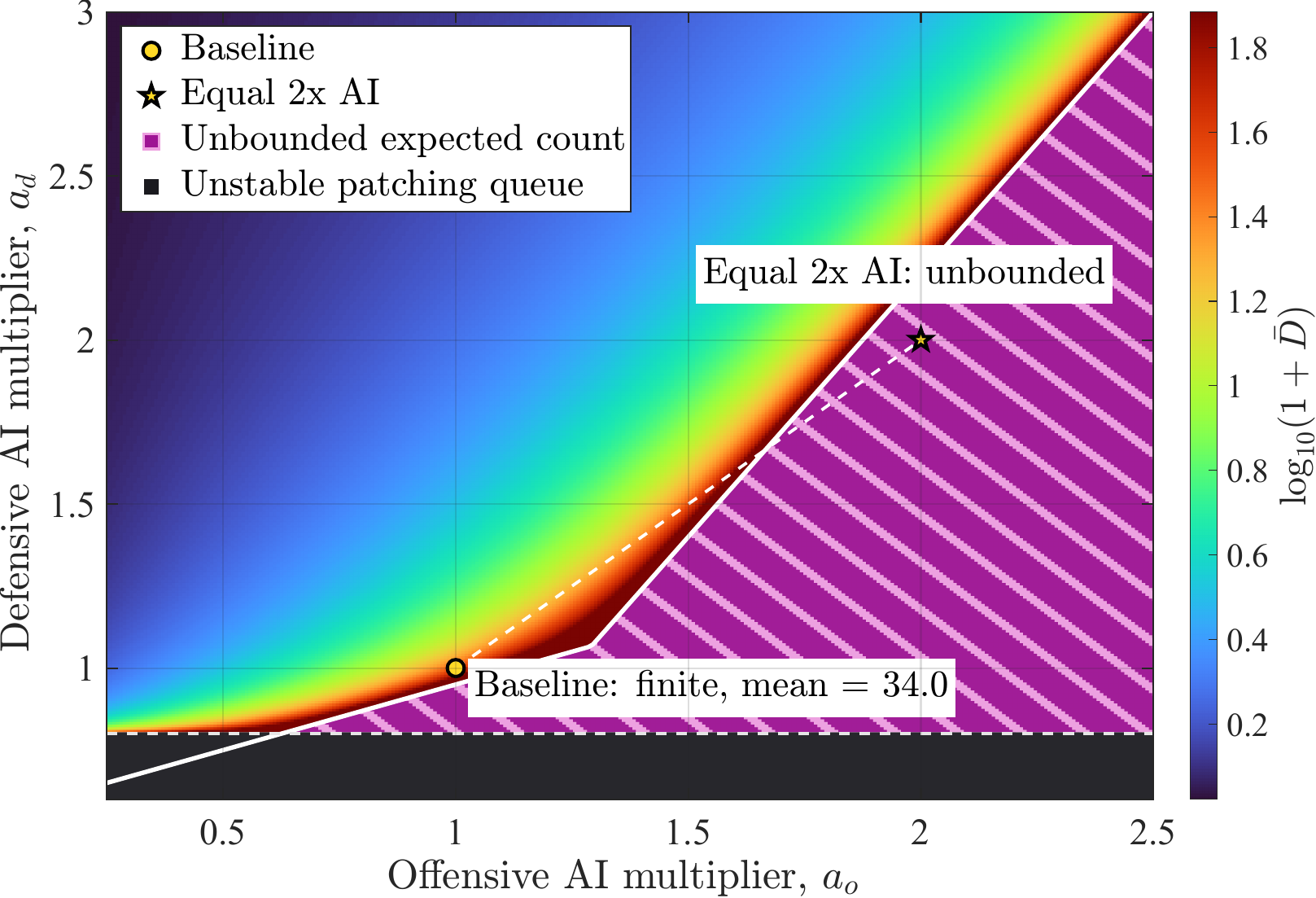}
	\caption{Expected cumulative number of exploits under offensive and defensive AI scaling. Colors show $\log_{10}(1+\bar{D})$ in the finite region. Magenta hatching denotes an unbounded expectation, and dark shading denotes an unstable patching queue. The white curve marks the phase boundary. The markers identify the baseline and equal twofold scaling scenarios.}
	\label{fig:ai-dbar-heatmap}
    \vspace*{-0.4cm}
\end{figure}

We now investigate the effects of AI on offensive and defensive cybersecurity capabilities for the BDI model. We represent their respective levels of amplification by the multipliers \(a_o\) and \(a_d\), respectively. Offensive AI increases both the per-attacker attraction rate and the independent attacker immigration rate. Defensive AI increases both the vulnerability discovery rate and the patching rate. Accordingly, the scaled rates are defined as
\begin{align}
    \lambda = a_o\lambda_{\mathrm{b}}, \
    \lambda_0 = \delta_a= a_o\lambda_{0,\mathrm{b}}, \
    \delta_d = a_d\delta_{d,\mathrm{b}}, \
    \mu = a_d\mu_{\mathrm{b}}.
    \label{eq:ai-scaling}
\end{align}
The vulnerability arrival rate $\gamma$ and the per-attacker exploit rate $\theta$ are held fixed so that the effects of offensive and defensive AI scaling can be isolated. For each pair $(a_o,a_d)$, we evaluate the expected cumulative number of exploits using \eqref{eq:barD}. We define the baseline as the reference scenario without AI amplification, for which $a_o=a_d=1$. For this scenario, we set $\gamma=0.8$, $\mu_{\mathrm{b}}=1$, $\delta_{d,\mathrm{b}}=0.25$, $\lambda_{\mathrm{b}}=0.4$, and $\theta=0.25$. The baseline independent attacker immigration rate is set to $\lambda_{0,\mathrm{b}}=0.4$.

Fig.~\ref{fig:ai-dbar-heatmap} shows that the baseline system lies in the finite region. At this point, the net attacker growth rate is $0.15$, while the discovery rate and effective patching rate are $0.25$ and $0.20$, respectively. The effective patching rate is therefore the limiting defensive mechanism, but it remains faster than the attacker growth. Substitution into \eqref{eq:barD} gives an expected cumulative exploit count of $\bar D=34$. The equal twofold scaling example produces a different outcome. Doubling the offensive capability increases the net attacker growth rate to $0.55$. Doubling the defensive capability increases the discovery rate to $0.50$ and the effective patching rate to $1.20$. Patching is then sufficiently fast, but discovery becomes the bottleneck. Because the attacker growth exceeds the discovery rate, the expected cumulative exploit count is unbounded. Thus, equal proportional improvements on both sides do not necessarily leave the system in the same operating regime.

The geometry of the white phase boundary explains why the two scenarios fall in different operating regimes. Its gently sloped left segment corresponds to the effective patching constraint and is the relevant boundary for the baseline point, whereas its steeper right segment corresponds to the vulnerability discovery constraint and bounds the equal twofold AI point. The bend near $(a_o,a_d)=(1.29,1.07)$ marks where the two defensive rates are equal and the bottleneck switches between them. Thus, whether the expected cumulative number of exploits remains finite is determined by the slower defensive mechanism, and improving the faster mechanism alone does not change the operating regime. Moreover, the expected number of exploits rises sharply near the boundary, thus a system can remain in the finite region while still being highly sensitive to parameter uncertainty or temporary attacker surges.

\section{Conclusion}

We introduce a queue of branching processes that couples vulnerability lifetimes with the evolution of attacker activity under both BDI and RD models. Our analysis shows that the expected number of exploits is not necessarily monotone in the exploit rate, since increasing the exploit rate can reduce the active attacker population sufficiently to offset its direct effect on exploit activity. Under the BDI model, attacker growth must remain below both the vulnerability discovery rate and the effective patching rate. We show that, under a fixed defensive resource budget, balancing these two defensive mechanisms minimizes the expected number of exploits and expands the range of attacker growth rates the system can sustain.

More broadly, the results show that the security of a dynamic attack surface is governed by the relative time scales of attack, discovery, and patching, rather than by the speed of any one mechanism in isolation. The slower defensive mechanism determines the effective bottleneck, and improving the faster mechanism alone may provide little benefit if the other cannot keep pace with attacker growth. This also changes how the impact of AI should be viewed: AI acts as an amplifier of the underlying system dynamics, accelerating both offensive and defensive processes. Equal proportional amplification on the two sides does not, in general, preserve the operating regime, since acceleration can shift the defensive bottleneck from one mechanism to another. In our numerical example, faster patching under AI amplification leaves vulnerability discovery as the limiting mechanism and drives the system from a finite-exploitation regime into one with an unbounded exploit count. Moreover, the rapid increase in expected exploitation near the phase boundary suggests that systems operating close to this boundary may be highly sensitive to small variations in operational rates. These results highlight the importance of jointly managing discovery and patching resources as the dynamics of both attack and defense continue to accelerate.

\bibliographystyle{IEEEtran}
\bibliography{bibl}

\end{document}